\documentclass[sigconf,]{acmart}
\AtBeginDocument{%
  \providecommand\BibTeX{{%
    \normalfont B\kern-0.5em{\scshape i\kern-0.25em b}\kern-0.8em\TeX}}}

\copyrightyear{2020}
\acmYear{2020}
\setcopyright{rightsretained}
\acmConference[CI2020]{10th International Conference on Climate
Informatics}{September 22--25, 2020}{virtual, United Kingdom}
\acmBooktitle{10th International Conference on Climate Informatics
(CI2020), September 22--25, 2020, virtual, United
Kingdom}\acmDOI{10.1145/3429309.3429315}
\acmISBN{978-1-4503-8848-1/20/09}

\begin{document}

\title{Spatio-temporal evolution of global surface temperature distributions}

\author{Federico Amato}
\orcid{0000-0002-5886-9038}
\affiliation{%
  \institution{University of Lausanne \\ Institute of Earth Surface Dynamics}
  \streetaddress{UNIL-Mouline}
  \city{Lausanne}
  \state{}
  \country{Switzerland}
  \postcode{CH-1015}
}
\email{federico.amato@unil.ch}

\author{Fabian Guignard}
\affiliation{%
  \institution{University of Lausanne \\ Institute of Earth Surface Dynamics}
  \streetaddress{UNIL-Mouline}
  \city{Lausanne}
  \state{}
  \country{Switzerland}
  \postcode{CH-1015}
}
\email{fabian.guignard@unil.ch}

\author{Vincent Humphrey}
\affiliation{%
 \institution{California Institute of Technology \\ Division of Geological and Planetary Sciences}
 \streetaddress{Rono-Hills}
 \city{Pasadena}
 \state{California}
 \country{USA}}
\email{vincent.humphrey@caltech.edu}

\author{Mikhail Kanevski}
\affiliation{%
  \institution{University of Lausanne \\ Institute of Earth Surface Dynamics}
  \streetaddress{UNIL-Mouline}
  \city{Lausanne}
  \state{}
  \country{Switzerland}
  \postcode{CH-1015}
}
\email{mikhail.kanevski@unil.ch}

\renewcommand{\shortauthors}{Amato, et al.}

\begin{abstract}
Climate is known for being characterised by strong non-linearity and chaotic behaviour. Nevertheless, few studies in climate science adopt statistical methods specifically designed for non-stationary or non-linear systems. Here we show how the use of statistical methods from Information Theory can describe the non-stationary behaviour of climate fields, unveiling spatial and temporal patterns that may otherwise be difficult to recognize. We study the maximum temperature at two meters above ground using the NCEP CDAS1 daily reanalysis data, with a spatial resolution of $2.5^{\circ}$ by $2.5^{\circ}$ and covering the time period from 1 January 1948 to 30 November 2018. The spatial and temporal evolution of the temperature time series are retrieved using the Fisher Information Measure, which quantifies the information in a signal, and the Shannon Entropy Power, which is a measure of its uncertainty --- or unpredictability. The results describe the temporal behaviour of the analysed variable. Our findings suggest that tropical and temperate zones are now characterized by higher levels of entropy. Finally, Fisher-Shannon Complexity is introduced and applied to study the evolution of the daily maximum surface temperature distributions.
\end{abstract}




\maketitle

\section{Motivation}
In the context of anthropogenic greenhouse gas emissions, global climate has been the focus of extensive research over the last decades. Countries which have adhered to the 2016 Paris Agreement aim to limit the increase in global average temperature below $1.5 ^\circ C$ and mitigate the risks and impacts of climate change \cite{hulme20161}. Moreover, climate change is also a primary concern in the context of the Sustainable Development Goals (SDGs) \cite{assembly2015sustainable}. In the United Nations agenda, climate influence on sustainable development is closely related not only to environmental issues but also to the social and economic dimension of the SDGs. Similarly, the Intergovernmental Panel on Climate Change (IPCC) has investigated the interaction between sustainable development, poverty eradication and ethics and equity \cite{intergovernmental2018global}. Many of these interactions are related to the fact that warming has significant spatial and temporal patterns that could affect some regions more than others. As an example many locations, especially in the Northern Hemisphere at the mid-latitudes, are experiencing regional warming that is more than double the global average \cite{collins2013long}.

Because climate is characterised by strong nonlinearity and chaotic behaviour, many studies in climate science cannot rely on statistical methods valid only for stationary or linear systems \cite{franzke2014warming, ribes2020describing}. It has already been shown that warming trends are characterised by strong non-linearities, with an acceleration in the increase of temperatures since 1980 \cite{ji2014evolution}.

In the present research we attempt to further investigate the complex nature of surface temperature, showing how statistical methods from Information Theory can be used to describe the non-stationary behaviour of such phenomena. We study the maximum temperature at two meters above ground using the NCEP CDAS1 daily reanalysis data, with a spatial resolution of $2.5^{\circ}$ by $2.5^{\circ}$ and covering the time period from 1 January 1948 to 30 November 2018 \cite{kalnay1996ncep}. For each spatial location we track through a sliding window the evolution of the corresponding temperature time series using the Fisher Information Measure (FIM) \cite{Fisher1925}, which is a powerful tool to identify the behaviour of dynamical systems, and the Shannon Entropy Power (SEP) \cite{Shannon1948}, which can be used to quantify the uncertainty --- or the disorder --- of a signal. We show how these measures can be used to retrieve the spatial and temporal changes of the investigated time series distributions, highlighting the complex behaviour of the temperature spatio-temporal field.

\section{Methods}

This section provides details concerning the estimation of FIM and SEP. Briefly, FIM is used to measure the information in a given signal $x$, while SEP quantifies its degree of disorder and predictability. To unveil the spatial structures of the spatio-temporal values of FIM and SEP computed, an Empirical Orthogonal Function (EOF) decomposition will also be performed on the data of the two estimations. EOF analysis is the spatio-temporal version of principal component analysis. Through this approach, data are decomposed into orthogonal basis functions defined by spatial eigenvectors --- the Empirical Orthogonal Functions (EOFs) ---  and temporal coefficients described by the principal components (PC) time series.

\subsection*{Fisher Information Measure and Shannon Entropy Power}

\begin{figure*}[hbt!]
\begin{center}
\includegraphics[width=1\linewidth]{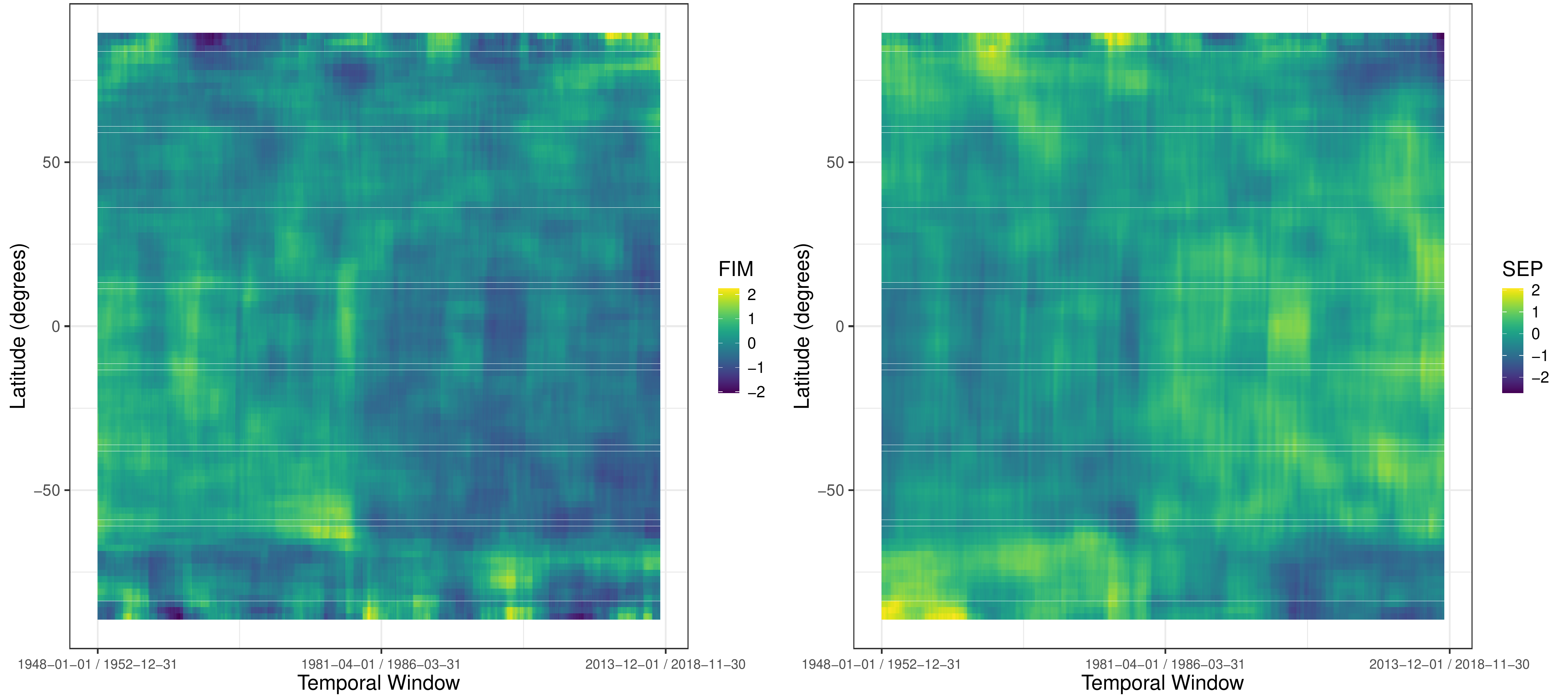}
\caption{\textbf{Hovm\"{o}ller plots for the latitude for both FIM and SEP.} The figures represent the mean values of the two measures over land and oceans. Data have been standardized using a z-score tranformation.}
\label{fig:hovmoller}
\end{center}
\end{figure*}

SEP and FIM are summary quantities to partially describe a Probability Density Function (PDF). Let $X$ be an univariate continuous random variable following a PDF  $f(x)$. The SEP of $X$ is defined as $N_X$ through the relationship \cite{Dembo1991}   
\begin{equation}\label{N}
N_X = \frac{1}{2\pi e} e^{2H_X}, \qquad H_X = \mathbb E \left[-\log f(x)\right].
\end{equation}
The SEP is a strictly increasing transformation of $H_X$, which is the differential entropy of $X$ \cite{Cover2006}.
The FIM of $X$, noted $I_X$, is defined as \cite{Dembo1991}. 
\begin{equation}\label{I}
    I_X=\mathbb E \left[\left(\frac{\partial}{\partial x}\log f(X)\right)^2\right].
\end{equation}

SEP and FIM can be jointly visualized into the Fisher-Shannon Information Plane (FSIP) to study the PDF of $X$ \cite{Vignat2003, guignard2019advanced}. It can be shown that $N_X \cdot I_X \geq 1$, with the equality satisfied if and only if $X$ is a Gaussian random variable \cite{Dembo1991}. Hence, the only reachable points in the FSIP belong to the set 
\[
\mathcal D = \left\{(N_X, I_X) \in \mathbb R^2 \;\middle|\;
\begin{aligned}
&N_X >0, \, I_X>0, \\ & N_X \cdot I_X \geq 1
\end{aligned} \right\}.
\]
The quantity $C_X = N_X \cdot I_X$, which is called Fisher-Shannon Complexity (FSC), is sometimes used as a statistical complexity measure \cite{Angulo2008, Esquivel2010}. 

It has been shown that the FSC can be interpreted as a measure of non-Gausiannity of $X$ \cite{guignard2019advanced}.
Thus, the boundary of $\mathcal D$ is reached if and only if $X$ has a unitary FSC, in which case it is a Gaussian random variable.

In order to estimate SEP and FIM from data, $f(x)$ and its derivative $f'(x)$ were replaced by their kernel density estimators (KDE) in the integral forms of \eqref{N} and \eqref{I}, \cite{Bhattacharya1967, Dmitriev1973, PrakasaRao1983, Gyorfi1987, Joe1989}.
To this aim, given $n$ independent realizations $\{x_1, \dots, x_n\}$ of $X$, the PDF $f(x)$ is approximated as  \cite{Wand1994}

\begin{equation}\label{KDE}
    \hat f_h(x) = \frac{1}{nh}\sum_{i=1}^n K\left( \frac{x-x_i}{h} \right),
\end{equation}
where $h$ is a bandwidth parameter and $K(\cdot)$ is a kernel assumed to be a unimodal probability density function symmetric around zero and having integral over $\mathbb R$ equal to 1. By using a Gaussian kernel with zero mean and unit variance, the estimator \eqref{KDE} assumes the form
\begin{equation}\label{KDEGauss}
    \hat f_h(x) = \frac{1}{nh\sqrt{2\pi}}\sum_{i=1}^n e^{-\frac{1}{2}\left( \frac{x-x_i}{h} \right)^2}.
\end{equation}
By deriving $\hat f_h(x)$ one can estimate the PDF derivative $f'(x)$ as
\begin{equation}\label{KDEderivGauss}
    \hat f'_h(x) = \frac{1}{nh^3 \sqrt{2\pi}}\sum_{i=1}^n (x-x_i)e^{-\frac{1}{2}\left( \frac{x-x_i}{h} \right)^2}.
\end{equation}
By plugging estimates \eqref{KDEGauss} and \eqref{KDEderivGauss} into the equations \eqref{N} and \eqref{I}, the latter assumes the form
\[
\hat N_X = \frac{1}{2\pi e}e^{-2\int \hat f_h(x) \log \hat f_h(x) \, dx},
\]
and
\[
\hat I_X = \int \frac{\left(f'_{h}(x)\right)^2}{f_h(x)} \, dx.
\]
The FSC is estimated by multiplying $\hat N_X$ by $\hat I_X$.
The estimates are sensitive to the choice of a proper bandwidth. Here, this parameter is selected using the Sheather-Jones direct plug-in method \cite{Sheather1991}, which approximates the optimal bandwidth with respect to the Asymptotic Mean Integrated Squared Error of $\hat f_h$. Operationally, the non-parametric estimation of the SEP, FIM and FSC are obtained with the FiShPy package \cite{guignard2019advanced}. 

\subsection*{Empirical Orthogonal Functions}

EOF analysis is an extremely popular approach to study the variability in a geophysical field
of interest \cite{hannachi2007empirical}. EOF analysis can be seen as an application of Principal Component Analysis in the case where data is a multivariate spatially indexed vector with multiple samples over time \cite{lorenz1956empirical, jolliffe2016principal}. 

Let  $\textbf{X}_{t_j}=\big( X(\textbf{s}_1;t_j),\dots, X(\textbf{s}_m;t_j)\big)' \in \mathbb R^m$ be observations for the spatial locations $\{\textbf{s}_i: i = 1,\dots, m\}$ and times $\{t_j: j = 1,\dots,T\}$. The empirical spatial mean for all location is given by
\begin{equation*}
    \widehat{\boldsymbol{\mu}}=\frac{1}{T}\sum_{j=1}^T \mathbf{X}_{t_j}.
\end{equation*}
The empirical spatial covariance matrix \cite{wikle2019spatio} can then be computed as
\begin{equation*}
    \widehat{\mathbf{C}} = \frac{1}{T} \sum_{j=1}^T (\mathbf{X}_{t_j} - \widehat{\boldsymbol{\mu}}) (\mathbf{X}_{t_j} - \widehat{\boldsymbol{\mu}})'.
\end{equation*}

As this real matrix is symmetric and non-negative definite, it can be spectrally decomposed as
\begin{equation*}
    \widehat{\mathbf{C}} = \boldsymbol\Phi \boldsymbol\Lambda \boldsymbol\Phi'
\end{equation*}
where $\boldsymbol\Lambda = \textrm{diag}(\lambda_1,\dots, \lambda_m)$ is the diagonal matrix of the non-negative eigenvalues decreasing down the diagonal, and $\boldsymbol\Phi=(\boldsymbol\phi_1,\dots, \boldsymbol\phi_m)$ is the matrix of the corresponding spatially indexed eigenvectors $\boldsymbol\phi_k = (\phi_k(\mathbf{s}_1), \dots, \phi_k(\mathbf{s}_m))'$, for $k=1,\dots,m$, also called EOFs. The EOFs form a  discrete orthonormal basis
\cite{cressie2015statistics}. The $k$-th PC time series --- which is the time series of coefficient of the corresponding EOF, or equivalently the contribution of the $k$-th spatial basis at time $t_j$ --- is then given by $a_k(t_j)=\boldsymbol\phi_k'\mathbf{X}_{t_j}$ \cite{monahan2009empirical}.

\section{Results}

\subsection*{Temporal changes of temperature distributions}

\begin{figure*}[ht!]
\begin{center}
\includegraphics[width=1\linewidth]{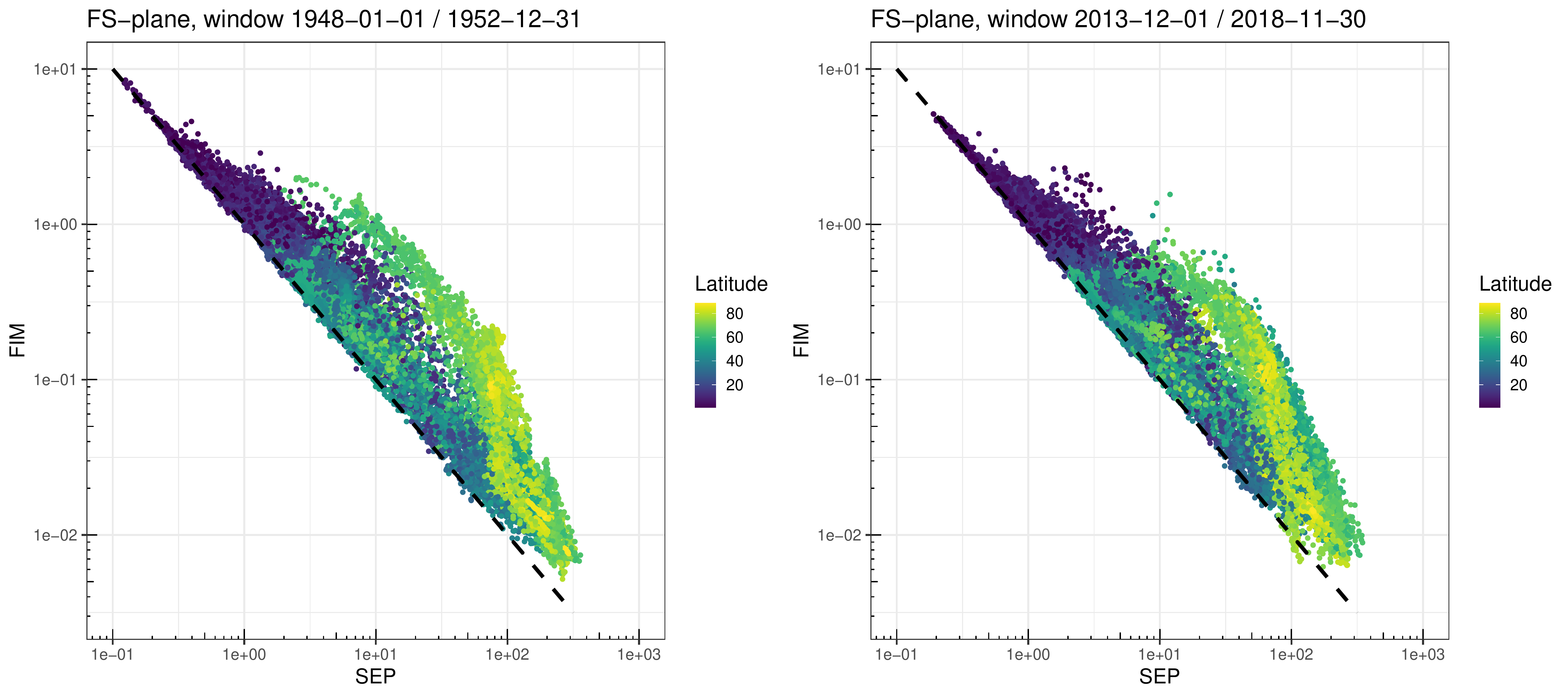}
\caption{\textbf{Fisher-Shannon Information Plane.} The FSIP constructed with the estimated values of FIM and SEP for the first (left) and last (right) sliding windows. Each point corresponds to a spatial location, while the dashed line represents the theoretical Gaussian limit. Color indicates the absolute value of the latitude of the points.}
\label{fig:fsplane}
\end{center}
\end{figure*}

To study the NCEP CDAS1 daily data, we first consider separately the temperature time series of the single spatial locations. For each of them, we computed time-dependent measures of SEP and FIM using a sliding temporal window of width equal to five years and a sliding factor equal to one month. 

To make spatial and temporal variabilities comparable, the obtained measures for each location have been standardized with a z-score transformation. The spatial distribution of FIM and SEP shown in Figure \ref{fig:hovmoller} highlights two different behaviour. In the period from 1948 to 1979 the latitudes higher than $60^{\circ}$ and lower than $-60^{\circ}$ are characterized by high levels of SEP and low levels of FIM. The situation has completely reversed starting from 1979, and in a faster way for the latitudes higher than $60^{\circ}$, so that tropical and temperate zones are now characterized by higher levels of entropy. The northern mid-latitude seems to have slightly stronger growth of SEP. 

The FSIP is used for a simultaneous analysis of FIM and SEP to track the non-stationary behaviour of the temperature time series \cite{Vignat2003}. Figure \ref{fig:fsplane} shows the FSIP of the non z-scored data for the first sliding window (from 1948/01/01 to 1952/12/31) and for the last one (from 2013/12/01 to 2018/11/30). The Figure highlights how the characterization of points in the FSIP is strongly linked to their latitude. The latitudes higher than $60^{\circ}$ and lower than $-60^{\circ}$ are easily recognizable for having a behaviour more distant from the one expected in the Gaussian case. Moreover, it is possible to observe a significant movement of these points from the first to the last window. The displacements of the points in the plane can be interpreted as a change of the distribution of temperature. 

\begin{figure*}[ht]
\begin{center}
\vspace{10mm}
\includegraphics[width=1\linewidth]{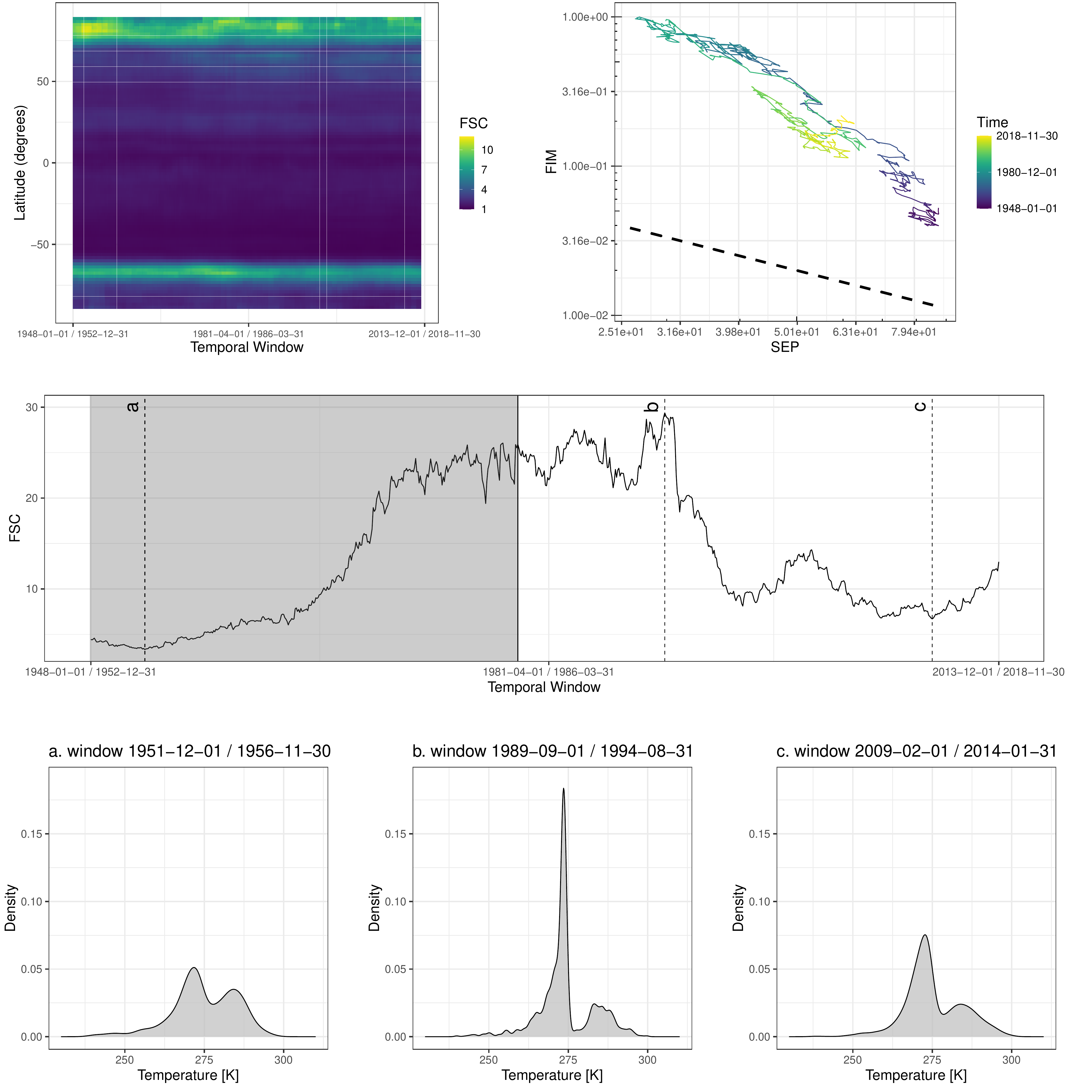}
\caption{\textbf{Fisher-Shannon Complexity.} Top left: Hovm\"{o}ller latitudinal plot for FSC. Top right: Trajectory in the FSIP of the point represented by a red triangle in Figure \ref{fig:eofs} and having Longitude =232.0 and Latitude = 56.2. The dashed line represents the theoretical Gaussian limit. Middle: FSC of the same point over time. The grey region covers the temporal period from 1948-01-01 to 1979-01-01. Bottom row: Distributions of the temperature measurements in the temporal windows resulting into the FSC points highlighted with a vertical dashed line in the previous plot.}
\label{fig:fsc}

\end{center}
\end{figure*}

FSC is analyzed in Figure \ref{fig:fsc}. The Hovm\"{o}ller plot highlights how the equatorial areas exhibit a FSC fluctuating around values not far from 1, implying temperature distributions almost Gaussian. Latitudes higher than $60^{\circ}$ and lower than $-60^{\circ}$ show a reduction of the FSC over time. 
Figure \ref{fig:fsc} also investigates the FSC of the point having Longitude = 232.0 and Latitude = 56.2, located in the western Canada. 
This is one of the point experiencing the greater overall change in the measured complexity over time among all the investigated spatial locations. The distribution changes of temperature at this location is clearly observed in the FSIP. The trajectory defined in the plane highlights different stages in its behaviour, which is also recognizable in the timeseries of the FSC. The latter show decreasing complexity values since September 1989. FSC can be used as a clear indicator of a changing pattern in the distribution of data. As an example, three distributions are shown, computed on the temporal windows from 1951-12-01 to 1956-11-0, from 1989-09-01 to 1994-08-31 and from 2009-02-01 to 2014-01-31, respectively.
All the distributions exhibit bimodality, where the modes are determined by the seasonals variabilities. The first distribution, associated with a FSC of 3.32, is actually not far from being a mixture of two Gaussian distributions. 
Indeed, it can be empirically observed that for a mixture of two Gaussian distributions having the same variance, the FSC goes to 4 when the modes move away from each other. Differently, the second distribution plot, corresponding to an FSC of 29.37, is dominated by the mode at 273K. This is still true in the last plot, corresponding to FSC of 6.67, although since September 1989 a reduction of the complexity has been registered, indicating a behaviour closer to Gaussian. 

\subsection*{Regional behaviour of temperature distributions}

\begin{figure*}[ht!]
\begin{center}
\includegraphics[width=1\linewidth]{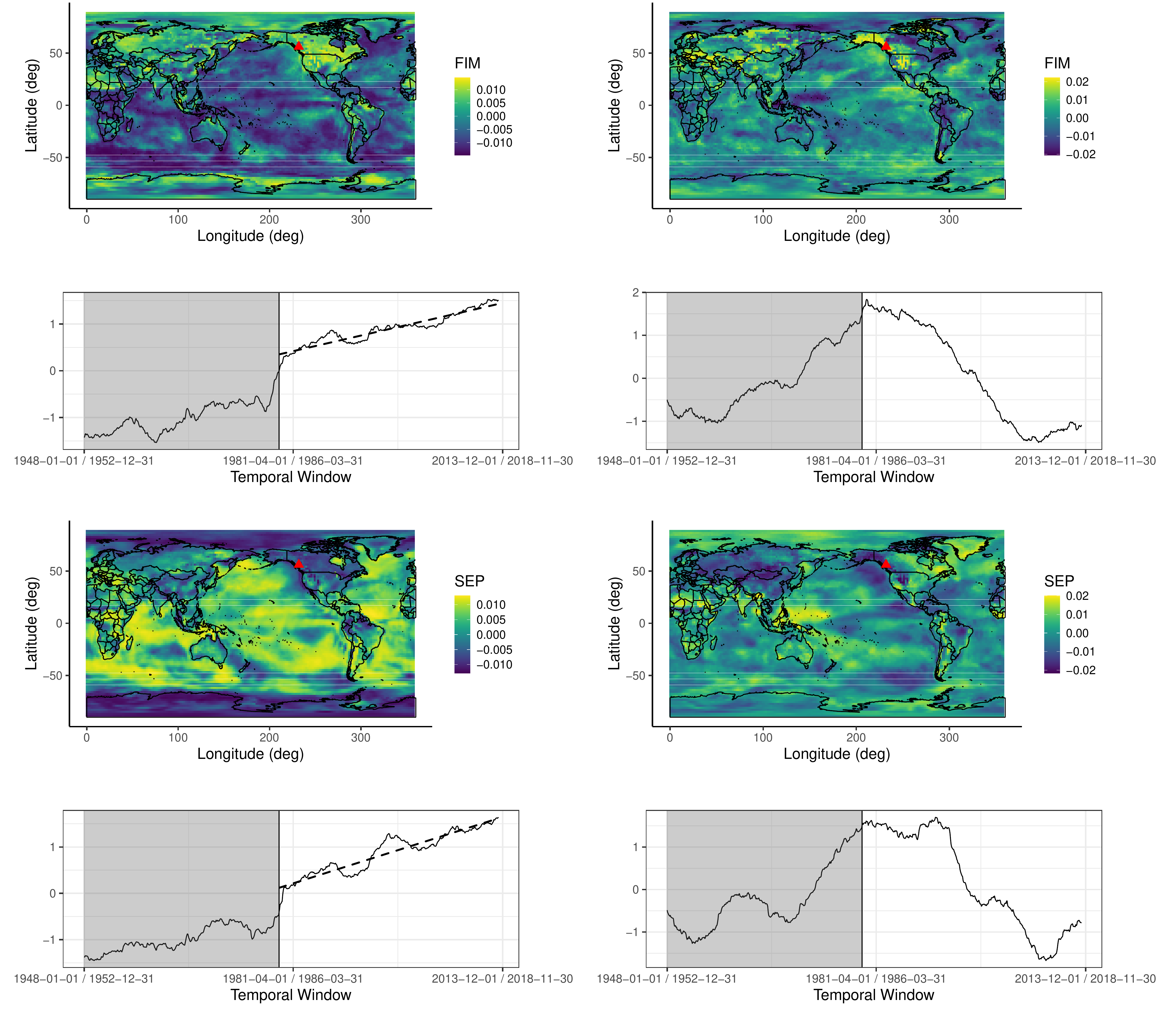}
\caption{\textbf{Empirical Orthogonal Function decomposition.} First two EOFs and corresponding PC time series for the FIM (top) and the SEP (bottom). The red triangle in the maps corresponds to the point explored in Figure \ref{fig:fsc}. The grey regions in the PC time series plots cover the temporal period from 1948-01-01 to 1979-01-01. For the PC time series corresponding to the first components, a regression line is fitted to the data from January 1979 and reported with a dashed line.}
\label{fig:eofs}
\end{center}
\end{figure*}

To further improve the understanding of the spatial patterns of the variability in the spatio-temporal values of the estimated SEP and FIM, EOF decompositions were performed. Figure \ref{fig:eofs} shows the first two EOFs and the corresponding PC time series for both SEP and FIM estimates.

The first EOFs reproduce the pattern of reduction of SEP and growth of FIM in the extreme northern and southern latitudes. The first EOFs also show a drastic difference between oceans and lands surfaces, with the former generally characterized by significant increases of SEP in the period after 1979. This may be related to the fact that the land surface experiences a stronger warming than the ocean surface, in relation to the higher evaporation and heat storage capacity of the oceans \cite{guemas2013retrospective}. It might also be related to changes in the atmospheric circulation (e.g. the  North Atlantic oscillation and the related arctic oscillation). This pattern might also reflect some spatial properties of the observing system. Because weather stations  and atmospheric soundings are typically land-based, the atmospheric reanalysis assimilates much more observations over land than over the ocean, particularly until the first satellite observations became available in 1979. The PC time series corresponding to the first EOFs show how the data are increasingly projecting on these spatial patterns starting from 1979. 

The second EOFs for SEP and FIM highlight interesting spatial patterns, once again related to significant changes starting from the late eighties as it shown by the corresponding PC time series. The most evident spatial trends concern the continental interiors of Asia, the western North America and central Brazil. These regions are described in the second EOFs by growing values of FIM up to the late eighties, with a drastic change from that period onward.  

\section{Conclusions}

This paper discussed how statistical methods derived from Information Theory could be used to investigate the properties of a climate field, specifically the surface air temperature. We found that the three measures applied, FIM, SEP and FSC, could provide meaningful insight about the global and local properties of the mentioned time series. Specifically, we were able to recover spatio-temporal structures in the data, depicting behaviours that, otherwise, would have been difficult to highlight.
Indeed, the results presented in the previous section underpin the capability of the applied measures to detect the degree to which observational products, such as the reanalyses ones, are affected by step changes in the underlying observing system. The detection of such changes is relevant, as a lack of their documentation may lead scientist to misleading conclusions regarding, for instance, climate trends. An example is the detection of two phases in the reanalysis dataset studied in this paper. The two are clearly separated by a behavioural change in 1979 recognized by both FIM and SEP. Indeed, starting from 1980 with the introduction of the Operational Vertical Sounder (TVOS), satellite data became part of the reanalysis model. The PC time series in Figure \ref{fig:eofs} are particularly effective to visualize this changing point.  
Nonetheless, even just considering data from 1979 it is possible to clearly identify spatial and temporal patterns in the two measures computed - as shown in Figure \ref{fig:eofs} by the regression lines on the PC time series of the first components of SEP ($R^2$ = 0.889) and FIM ($R^2$ = 0.902).

Warming is not a smooth monotonous process, and the chaotic nature of climate systems reduces the possibility of performing reliable forecasts of future temperature scenarios \cite{sevellec2018novel}. Previous studies shown that entropy can be used to spatially identify unpredictability patterns in surface temperature data \cite{arizmendi2017identifying}. The joint use of SEP and FIM highlighted how the spatial and temporal evolution of this predictability level may not be constant over the entire globe and along time, stressing how the increase of observational data could not always lead to a decrease of such unpredictability \cite{zidek2003uncertainty}. However, the effect of the distribution changes along time on the medium to long term forecasts will have to be further investigated.

Future work will also have to focus on the study of the trend recognized in SEP and FIM since 1979. It is not possible to infer the causal factors inducing this trend using only the NCEP CDAS1 data, as it is not possible to distinguish the effects induced by the growing number of available observations used by the model and those caused by a changing climate. Hence, future analysis will be based on factorial climate simulations from the Coupled Model Intercomparison Project, to compare the information/entropy behaviour of pre-industrial control simulations versus historical simulations including the effects of land use change and greenhouse gas emissions. This could recognize specific patterns due to e.g. GHG emissions. 

\begin{acks}
The research presented in this paper was partly supported by the National Research Program 75 "Big Data" (PNR75, project No. 167285 "HyEnergy") of the Swiss National Science Foundation (SNSF). V.H. is supported by the SNSF grant no. P400P2\_180784.
\end{acks}

\bibliographystyle{ACM-Reference-Format}
\bibliography{sample-base}


\begin{thebibliography}{33}


\ifx \showCODEN    \undefined \def \showCODEN     #1{\unskip}     \fi
\ifx \showDOI      \undefined \def \showDOI       #1{#1}\fi
\ifx \showISBNx    \undefined \def \showISBNx     #1{\unskip}     \fi
\ifx \showISBNxiii \undefined \def \showISBNxiii  #1{\unskip}     \fi
\ifx \showISSN     \undefined \def \showISSN      #1{\unskip}     \fi
\ifx \showLCCN     \undefined \def \showLCCN      #1{\unskip}     \fi
\ifx \shownote     \undefined \def \shownote      #1{#1}          \fi
\ifx \showarticletitle \undefined \def \showarticletitle #1{#1}   \fi
\ifx \showURL      \undefined \def \showURL       {\relax}        \fi
\providecommand\bibfield[2]{#2}
\providecommand\bibinfo[2]{#2}
\providecommand\natexlab[1]{#1}
\providecommand\showeprint[2][]{arXiv:#2}

\bibitem[\protect\citeauthoryear{Angulo, Antolín, and Sen}{Angulo
  et~al\mbox{.}}{2008}]%
        {Angulo2008}
\bibfield{author}{\bibinfo{person}{J.C. Angulo}, \bibinfo{person}{J. Antolín},
  {and} \bibinfo{person}{K.D. Sen}.} \bibinfo{year}{2008}\natexlab{}.
\newblock \showarticletitle{{F}isher–{S}hannon plane and statistical
  complexity of atoms}.
\newblock \bibinfo{journal}{\emph{Physics Letters A}} \bibinfo{volume}{372},
  \bibinfo{number}{5} (\bibinfo{year}{2008}), \bibinfo{pages}{670 -- 674}.
\newblock
\showISSN{0375-9601}
\urldef\tempurl%
\url{https://doi.org/10.1016/j.physleta.2007.07.077}
\showDOI{\tempurl}


\bibitem[\protect\citeauthoryear{Arizmendi, Barreiro, and Masoller}{Arizmendi
  et~al\mbox{.}}{2017}]%
        {arizmendi2017identifying}
\bibfield{author}{\bibinfo{person}{Fernando Arizmendi},
  \bibinfo{person}{Marcelo Barreiro}, {and} \bibinfo{person}{Cristina
  Masoller}.} \bibinfo{year}{2017}\natexlab{}.
\newblock \showarticletitle{Identifying large-scale patterns of
  unpredictability and response to insolation in atmospheric data}.
\newblock \bibinfo{journal}{\emph{Scientific reports}}  \bibinfo{volume}{7}
  (\bibinfo{year}{2017}), \bibinfo{pages}{45676}.
\newblock


\bibitem[\protect\citeauthoryear{Bhattacharya}{Bhattacharya}{1967}]%
        {Bhattacharya1967}
\bibfield{author}{\bibinfo{person}{P.~K. Bhattacharya}.}
  \bibinfo{year}{1967}\natexlab{}.
\newblock \showarticletitle{Estimation of a Probability Density Function and
  Its Derivatives}.
\newblock \bibinfo{journal}{\emph{The Indian Journal of Statistics, Series A
  (1961-2002)}} \bibinfo{volume}{29}, \bibinfo{number}{4}
  (\bibinfo{year}{1967}), \bibinfo{pages}{373--382}.
\newblock


\bibitem[\protect\citeauthoryear{Collins, Knutti, Arblaster, Dufresne,
  Fichefet, Friedlingstein, Gao, Gutowski, Johns, Krinner,
  et~al\mbox{.}}{Collins et~al\mbox{.}}{2013}]%
        {collins2013long}
\bibfield{author}{\bibinfo{person}{Matthew Collins}, \bibinfo{person}{Reto
  Knutti}, \bibinfo{person}{Julie Arblaster}, \bibinfo{person}{Jean-Louis
  Dufresne}, \bibinfo{person}{Thierry Fichefet}, \bibinfo{person}{Pierre
  Friedlingstein}, \bibinfo{person}{Xuejie Gao}, \bibinfo{person}{William~J
  Gutowski}, \bibinfo{person}{Tim Johns}, \bibinfo{person}{Gerhard Krinner},
  {et~al\mbox{.}}} \bibinfo{year}{2013}\natexlab{}.
\newblock \showarticletitle{Long-term climate change: projections, commitments
  and irreversibility}.
\newblock In \bibinfo{booktitle}{\emph{Climate Change 2013-The Physical Science
  Basis: Contribution of Working Group I to the Fifth Assessment Report of the
  Intergovernmental Panel on Climate Change}}. \bibinfo{publisher}{Cambridge
  University Press}, \bibinfo{pages}{1029--1136}.
\newblock


\bibitem[\protect\citeauthoryear{Cover and Thomas}{Cover and Thomas}{2006}]%
        {Cover2006}
\bibfield{author}{\bibinfo{person}{Thomas~M. Cover} {and}
  \bibinfo{person}{Joy~A. Thomas}.} \bibinfo{year}{2006}\natexlab{}.
\newblock \bibinfo{booktitle}{\emph{Elements of Information Theory (Wiley
  Series in Telecommunications and Signal Processing)}}.
\newblock \bibinfo{publisher}{Wiley-Interscience}, \bibinfo{address}{New York,
  NY, USA}.
\newblock
\showISBNx{0471241954}


\bibitem[\protect\citeauthoryear{Cressie and Wikle}{Cressie and Wikle}{2015}]%
        {cressie2015statistics}
\bibfield{author}{\bibinfo{person}{Noel Cressie} {and}
  \bibinfo{person}{Christopher~K Wikle}.} \bibinfo{year}{2015}\natexlab{}.
\newblock \bibinfo{booktitle}{\emph{Statistics for spatio-temporal data}}.
\newblock \bibinfo{publisher}{John Wiley \& Sons}.
\newblock


\bibitem[\protect\citeauthoryear{Dembo, Cover, and Thomas}{Dembo
  et~al\mbox{.}}{1991}]%
        {Dembo1991}
\bibfield{author}{\bibinfo{person}{A. Dembo}, \bibinfo{person}{T.~M. Cover},
  {and} \bibinfo{person}{J.~A. Thomas}.} \bibinfo{year}{1991}\natexlab{}.
\newblock \showarticletitle{Information theoretic inequalities}.
\newblock \bibinfo{journal}{\emph{IEEE Transactions on Information Theory}}
  \bibinfo{volume}{37}, \bibinfo{number}{6} (\bibinfo{date}{Nov}
  \bibinfo{year}{1991}), \bibinfo{pages}{1501--1518}.
\newblock
\showISSN{0018-9448}
\urldef\tempurl%
\url{https://doi.org/10.1109/18.104312}
\showDOI{\tempurl}


\bibitem[\protect\citeauthoryear{Dmitriev and Tarasenko}{Dmitriev and
  Tarasenko}{1973}]%
        {Dmitriev1973}
\bibfield{author}{\bibinfo{person}{Y. Dmitriev} {and} \bibinfo{person}{F.
  Tarasenko}.} \bibinfo{year}{1973}\natexlab{}.
\newblock \showarticletitle{On the Estimation of Functionals of the Probability
  Density and Its Derivatives}.
\newblock \bibinfo{journal}{\emph{Theory of Probability \& Its Applications}}
  \bibinfo{volume}{18}, \bibinfo{number}{3} (\bibinfo{year}{1973}),
  \bibinfo{pages}{628--633}.
\newblock
\urldef\tempurl%
\url{https://doi.org/10.1137/1118083}
\showDOI{\tempurl}


\bibitem[\protect\citeauthoryear{Esquivel, Angulo, Antolín, Dehesa,
  López-Rosa, and Flores-Gallegos}{Esquivel et~al\mbox{.}}{2010}]%
        {Esquivel2010}
\bibfield{author}{\bibinfo{person}{Rodolfo~O. Esquivel},
  \bibinfo{person}{Juan~Carlos Angulo}, \bibinfo{person}{Juan Antolín},
  \bibinfo{person}{Jesús~S. Dehesa}, \bibinfo{person}{Sheila López-Rosa},
  {and} \bibinfo{person}{Nelson Flores-Gallegos}.}
  \bibinfo{year}{2010}\natexlab{}.
\newblock \showarticletitle{Analysis of complexity measures and information
  planes of selected molecules in position and momentum spaces}.
\newblock \bibinfo{journal}{\emph{Phys. Chem. Chem. Phys.}}
  \bibinfo{volume}{12} (\bibinfo{year}{2010}), \bibinfo{pages}{7108--7116}.
\newblock
Issue 26.
\urldef\tempurl%
\url{https://doi.org/10.1039/B927055H}
\showDOI{\tempurl}


\bibitem[\protect\citeauthoryear{Fisher}{Fisher}{1925}]%
        {Fisher1925}
\bibfield{author}{\bibinfo{person}{R.~A. Fisher}.}
  \bibinfo{year}{1925}\natexlab{}.
\newblock \showarticletitle{Theory of Statistical Estimation}.
\newblock \bibinfo{journal}{\emph{Mathematical Proceedings of the Cambridge
  Philosophical Society}} \bibinfo{volume}{22}, \bibinfo{number}{5}
  (\bibinfo{year}{1925}), \bibinfo{pages}{700–725}.
\newblock
\urldef\tempurl%
\url{https://doi.org/10.1017/S0305004100009580}
\showDOI{\tempurl}


\bibitem[\protect\citeauthoryear{Franzke}{Franzke}{2014}]%
        {franzke2014warming}
\bibfield{author}{\bibinfo{person}{Christian~LE Franzke}.}
  \bibinfo{year}{2014}\natexlab{}.
\newblock \showarticletitle{Warming trends: nonlinear climate change}.
\newblock \bibinfo{journal}{\emph{Nature Climate Change}} \bibinfo{volume}{4},
  \bibinfo{number}{6} (\bibinfo{year}{2014}), \bibinfo{pages}{423}.
\newblock


\bibitem[\protect\citeauthoryear{GA}{GA}{2015}]%
        {assembly2015sustainable}
\bibfield{author}{\bibinfo{person}{UN GA}.} \bibinfo{year}{2015}\natexlab{}.
\newblock \showarticletitle{Transforming our world: the 2030 Agenda for
  Sustainable Development}.
\newblock \bibinfo{journal}{\emph{Division for Sustainable Development Goals:
  New York, NY, USA}} (\bibinfo{year}{2015}).
\newblock


\bibitem[\protect\citeauthoryear{Guemas, Doblas-Reyes, Andreu-Burillo, and
  Asif}{Guemas et~al\mbox{.}}{2013}]%
        {guemas2013retrospective}
\bibfield{author}{\bibinfo{person}{Virginie Guemas},
  \bibinfo{person}{Francisco~J Doblas-Reyes}, \bibinfo{person}{Isabel
  Andreu-Burillo}, {and} \bibinfo{person}{Muhammad Asif}.}
  \bibinfo{year}{2013}\natexlab{}.
\newblock \showarticletitle{Retrospective prediction of the global warming
  slowdown in the past decade}.
\newblock \bibinfo{journal}{\emph{Nature Climate Change}} \bibinfo{volume}{3},
  \bibinfo{number}{7} (\bibinfo{year}{2013}), \bibinfo{pages}{649}.
\newblock


\bibitem[\protect\citeauthoryear{Guignard, Laib, Amato, and Kanevski}{Guignard
  et~al\mbox{.}}{2020}]%
        {guignard2019advanced}
\bibfield{author}{\bibinfo{person}{Fabian Guignard}, \bibinfo{person}{Mohamed
  Laib}, \bibinfo{person}{Federico Amato}, {and} \bibinfo{person}{Mikhail
  Kanevski}.} \bibinfo{year}{2020}\natexlab{}.
\newblock \showarticletitle{Advanced analysis of temporal data using
  {F}isher-{S}hannon information: theoretical development and application in
  geosciences}.
\newblock \bibinfo{journal}{\emph{Frontiers in Earth Science}}
  (\bibinfo{year}{2020}).
\newblock


\bibitem[\protect\citeauthoryear{Györfi and van~der Meulen}{Györfi and
  van~der Meulen}{1987}]%
        {Gyorfi1987}
\bibfield{author}{\bibinfo{person}{László Györfi} {and}
  \bibinfo{person}{Edward~C. van~der Meulen}.} \bibinfo{year}{1987}\natexlab{}.
\newblock \showarticletitle{Density-free convergence properties of various
  estimators of entropy}.
\newblock \bibinfo{journal}{\emph{Computational Statistics and Data Analysis}}
  \bibinfo{volume}{5}, \bibinfo{number}{4} (\bibinfo{year}{1987}),
  \bibinfo{pages}{425 -- 436}.
\newblock
\showISSN{0167-9473}
\urldef\tempurl%
\url{https://doi.org/10.1016/0167-9473(87)90065-X}
\showDOI{\tempurl}


\bibitem[\protect\citeauthoryear{Hannachi, Jolliffe, and Stephenson}{Hannachi
  et~al\mbox{.}}{2007}]%
        {hannachi2007empirical}
\bibfield{author}{\bibinfo{person}{A Hannachi}, \bibinfo{person}{IT Jolliffe},
  {and} \bibinfo{person}{DB Stephenson}.} \bibinfo{year}{2007}\natexlab{}.
\newblock \showarticletitle{Empirical orthogonal functions and related
  techniques in atmospheric science: A review}.
\newblock \bibinfo{journal}{\emph{International Journal of Climatology: A
  Journal of the Royal Meteorological Society}} \bibinfo{volume}{27},
  \bibinfo{number}{9} (\bibinfo{year}{2007}), \bibinfo{pages}{1119--1152}.
\newblock


\bibitem[\protect\citeauthoryear{Hulme}{Hulme}{2016}]%
        {hulme20161}
\bibfield{author}{\bibinfo{person}{Mike Hulme}.}
  \bibinfo{year}{2016}\natexlab{}.
\newblock \showarticletitle{1.5$^o{C}$ and climate research after the Paris
  Agreement}.
\newblock \bibinfo{journal}{\emph{Nature Climate Change}} \bibinfo{volume}{6},
  \bibinfo{number}{3} (\bibinfo{year}{2016}), \bibinfo{pages}{222}.
\newblock


\bibitem[\protect\citeauthoryear{IPCC}{IPCC}{2018}]%
        {intergovernmental2018global}
\bibfield{author}{\bibinfo{person}{IPCC}.} \bibinfo{year}{2018}\natexlab{}.
\newblock \bibinfo{booktitle}{\emph{Global Warming of 1.5$^oC$ : An IPCC
  Special Report on the Impacts of Global Warming of 1.5$^oC$ Above
  Pre-industrial Levels and Related Global Greenhouse Gas Emission Pathways, in
  the Context of Strengthening the Global Response to the Threat of Climate
  Change, Sustainable Development, and Efforts to Eradicate Poverty}}.
\newblock \bibinfo{publisher}{Intergovernmental Panel on Climate Change}.
\newblock


\bibitem[\protect\citeauthoryear{Ji, Wu, Huang, and Chassignet}{Ji
  et~al\mbox{.}}{2014}]%
        {ji2014evolution}
\bibfield{author}{\bibinfo{person}{Fei Ji}, \bibinfo{person}{Zhaohua Wu},
  \bibinfo{person}{Jianping Huang}, {and} \bibinfo{person}{Eric~P Chassignet}.}
  \bibinfo{year}{2014}\natexlab{}.
\newblock \showarticletitle{Evolution of land surface air temperature trend}.
\newblock \bibinfo{journal}{\emph{Nature Climate Change}} \bibinfo{volume}{4},
  \bibinfo{number}{6} (\bibinfo{year}{2014}), \bibinfo{pages}{462}.
\newblock


\bibitem[\protect\citeauthoryear{Joe}{Joe}{1989}]%
        {Joe1989}
\bibfield{author}{\bibinfo{person}{Harry Joe}.}
  \bibinfo{year}{1989}\natexlab{}.
\newblock \showarticletitle{Estimation of entropy and other functionals of a
  multivariate density}.
\newblock \bibinfo{journal}{\emph{Annals of the Institute of Statistical
  Mathematics}} \bibinfo{volume}{41}, \bibinfo{number}{4} (\bibinfo{date}{01
  Dec} \bibinfo{year}{1989}), \bibinfo{pages}{683--697}.
\newblock
\showISSN{1572-9052}
\urldef\tempurl%
\url{https://doi.org/10.1007/BF00057735}
\showDOI{\tempurl}


\bibitem[\protect\citeauthoryear{Jolliffe and Cadima}{Jolliffe and
  Cadima}{2016}]%
        {jolliffe2016principal}
\bibfield{author}{\bibinfo{person}{Ian~T Jolliffe} {and} \bibinfo{person}{Jorge
  Cadima}.} \bibinfo{year}{2016}\natexlab{}.
\newblock \showarticletitle{Principal component analysis: a review and recent
  developments}.
\newblock \bibinfo{journal}{\emph{Philosophical Transactions of the Royal
  Society A: Mathematical, Physical and Engineering Sciences}}
  \bibinfo{volume}{374}, \bibinfo{number}{2065} (\bibinfo{year}{2016}),
  \bibinfo{pages}{20150202}.
\newblock


\bibitem[\protect\citeauthoryear{Kalnay, Kanamitsu, Kistler, Collins, Deaven,
  Gandin, Iredell, Saha, White, Woollen, et~al\mbox{.}}{Kalnay
  et~al\mbox{.}}{1996}]%
        {kalnay1996ncep}
\bibfield{author}{\bibinfo{person}{Eugenia Kalnay}, \bibinfo{person}{Masao
  Kanamitsu}, \bibinfo{person}{Robert Kistler}, \bibinfo{person}{William
  Collins}, \bibinfo{person}{Dennis Deaven}, \bibinfo{person}{Lev Gandin},
  \bibinfo{person}{Mark Iredell}, \bibinfo{person}{Suranjana Saha},
  \bibinfo{person}{Glenn White}, \bibinfo{person}{John Woollen},
  {et~al\mbox{.}}} \bibinfo{year}{1996}\natexlab{}.
\newblock \showarticletitle{The NCEP/NCAR 40-year reanalysis project}.
\newblock \bibinfo{journal}{\emph{Bulletin of the American meteorological
  Society}} \bibinfo{volume}{77}, \bibinfo{number}{3} (\bibinfo{year}{1996}),
  \bibinfo{pages}{437--472}.
\newblock


\bibitem[\protect\citeauthoryear{Lorenz}{Lorenz}{1956}]%
        {lorenz1956empirical}
\bibfield{author}{\bibinfo{person}{Edward~N Lorenz}.}
  \bibinfo{year}{1956}\natexlab{}.
\newblock \showarticletitle{Empirical orthogonal functions and statistical
  weather prediction}.
\newblock \bibinfo{journal}{\emph{Massachusetts Institute of Technology,
  Department of Meteorology Cambridge}} (\bibinfo{year}{1956}).
\newblock


\bibitem[\protect\citeauthoryear{Monahan, Fyfe, Ambaum, Stephenson, and
  North}{Monahan et~al\mbox{.}}{2009}]%
        {monahan2009empirical}
\bibfield{author}{\bibinfo{person}{Adam~H Monahan}, \bibinfo{person}{John~C
  Fyfe}, \bibinfo{person}{Maarten~HP Ambaum}, \bibinfo{person}{David~B
  Stephenson}, {and} \bibinfo{person}{Gerald~R North}.}
  \bibinfo{year}{2009}\natexlab{}.
\newblock \showarticletitle{Empirical orthogonal functions: The medium is the
  message}.
\newblock \bibinfo{journal}{\emph{Journal of Climate}} \bibinfo{volume}{22},
  \bibinfo{number}{24} (\bibinfo{year}{2009}), \bibinfo{pages}{6501--6514}.
\newblock


\bibitem[\protect\citeauthoryear{Prakasa~Rao}{Prakasa~Rao}{1983}]%
        {PrakasaRao1983}
\bibfield{author}{\bibinfo{person}{B.L.S. Prakasa~Rao}.}
  \bibinfo{year}{1983}\natexlab{}.
\newblock \bibinfo{booktitle}{\emph{Nonparametric Functional Estimation}}.
\newblock \bibinfo{publisher}{Academic Press}.
\newblock


\bibitem[\protect\citeauthoryear{Ribes, Thao, and Cattiaux}{Ribes
  et~al\mbox{.}}{2020}]%
        {ribes2020describing}
\bibfield{author}{\bibinfo{person}{Aur{\'e}lien Ribes},
  \bibinfo{person}{Soulivanh Thao}, {and} \bibinfo{person}{Julien Cattiaux}.}
  \bibinfo{year}{2020}\natexlab{}.
\newblock \showarticletitle{Describing the relationship between a weather event
  and climate change: a new statistical approach}.
\newblock \bibinfo{journal}{\emph{Journal of Climate}} \bibinfo{number}{2020}
  (\bibinfo{year}{2020}).
\newblock


\bibitem[\protect\citeauthoryear{S{\'e}vellec and Drijfhout}{S{\'e}vellec and
  Drijfhout}{2018}]%
        {sevellec2018novel}
\bibfield{author}{\bibinfo{person}{Florian S{\'e}vellec} {and}
  \bibinfo{person}{Sybren~S Drijfhout}.} \bibinfo{year}{2018}\natexlab{}.
\newblock \showarticletitle{A novel probabilistic forecast system predicting
  anomalously warm 2018-2022 reinforcing the long-term global warming trend}.
\newblock \bibinfo{journal}{\emph{Nature communications}} \bibinfo{volume}{9},
  \bibinfo{number}{1} (\bibinfo{year}{2018}), \bibinfo{pages}{3024}.
\newblock


\bibitem[\protect\citeauthoryear{Shannon}{Shannon}{1948}]%
        {Shannon1948}
\bibfield{author}{\bibinfo{person}{C.~E. Shannon}.}
  \bibinfo{year}{1948}\natexlab{}.
\newblock \showarticletitle{A Mathematical Theory of Communication}.
\newblock \bibinfo{journal}{\emph{Bell System Technical Journal}}
  \bibinfo{volume}{27}, \bibinfo{number}{3} (\bibinfo{year}{1948}),
  \bibinfo{pages}{379--423}.
\newblock
\urldef\tempurl%
\url{https://doi.org/10.1002/j.1538-7305.1948.tb01338.x}
\showDOI{\tempurl}


\bibitem[\protect\citeauthoryear{Sheather and Jones}{Sheather and
  Jones}{1991}]%
        {Sheather1991}
\bibfield{author}{\bibinfo{person}{S.~J. Sheather} {and} \bibinfo{person}{M.~C.
  Jones}.} \bibinfo{year}{1991}\natexlab{}.
\newblock \showarticletitle{A Reliable Data-Based Bandwidth Selection Method
  for Kernel Density Estimation}.
\newblock \bibinfo{journal}{\emph{Journal of the Royal Statistical Society.
  Series B (Methodological)}} \bibinfo{volume}{53}, \bibinfo{number}{3}
  (\bibinfo{year}{1991}), \bibinfo{pages}{683--690}.
\newblock
\showISSN{00359246}


\bibitem[\protect\citeauthoryear{Vignat and Bercher}{Vignat and
  Bercher}{2003}]%
        {Vignat2003}
\bibfield{author}{\bibinfo{person}{C. Vignat} {and} \bibinfo{person}{J.-F.
  Bercher}.} \bibinfo{year}{2003}\natexlab{}.
\newblock \showarticletitle{Analysis of signals in the Fisher–Shannon
  information plane}.
\newblock \bibinfo{journal}{\emph{Physics Letters A}} \bibinfo{volume}{312},
  \bibinfo{number}{1} (\bibinfo{year}{2003}), \bibinfo{pages}{27 -- 33}.
\newblock
\showISSN{0375-9601}
\urldef\tempurl%
\url{https://doi.org/10.1016/S0375-9601(03)00570-X}
\showDOI{\tempurl}


\bibitem[\protect\citeauthoryear{Wand and Jones}{Wand and Jones}{1994}]%
        {Wand1994}
\bibfield{author}{\bibinfo{person}{M.P. Wand} {and} \bibinfo{person}{M.C.
  Jones}.} \bibinfo{year}{1994}\natexlab{}.
\newblock \bibinfo{booktitle}{\emph{Kernel Smoothing}}.
\newblock \bibinfo{publisher}{Taylor \& Francis}.
\newblock
\showISBNx{9780412552700}


\bibitem[\protect\citeauthoryear{Wikle, Zammit-Mangion, and Cressie}{Wikle
  et~al\mbox{.}}{2019}]%
        {wikle2019spatio}
\bibfield{author}{\bibinfo{person}{Christopher~K Wikle},
  \bibinfo{person}{Andrew Zammit-Mangion}, {and} \bibinfo{person}{Noel
  Cressie}.} \bibinfo{year}{2019}\natexlab{}.
\newblock \bibinfo{booktitle}{\emph{Spatio-temporal Statistics with R}}.
\newblock \bibinfo{publisher}{CRC Press}.
\newblock


\bibitem[\protect\citeauthoryear{Zidek and van Eeden}{Zidek and van
  Eeden}{2003}]%
        {zidek2003uncertainty}
\bibfield{author}{\bibinfo{person}{James~V Zidek} {and}
  \bibinfo{person}{Constance van Eeden}.} \bibinfo{year}{2003}\natexlab{}.
\newblock \showarticletitle{Uncertainty, entropy, variance and the effect of
  partial information}.
\newblock \bibinfo{journal}{\emph{Lecture Notes-Monograph Series}}
  (\bibinfo{year}{2003}), \bibinfo{pages}{155--167}.
\newblock


\end{thebibliography}

\end{document}